# An investigation into a wavelet accelerated gauge fixing algorithm

Terrence Draper and Craig McNeile[a]

[a]Department of Physics and Astronomy, University of Kentucky,
Lexington, KY 40506, USA

We introduce an acceleration algorithm for coulomb gauge fixing, using the compactly supported wavelets introduced by Daubechies. The algorithm is similar to Fourier acceleration. Our provisional numerical results for $SU(3)$ on $8^4$ lattices show that the acceleration based on the DAUB6 transform can reduce the number of iterations by a factor up to 3 over the unaccelerated algorithm. The reduction in iterations for Fourier acceleration is approximately a factor of 7.

## 1. INTRODUCTION

The performances of many lattice gauge theory algorithms are hampered by critical slowing down. The maximum stable step size of an algorithm is governed by the smallest length scale in the problem. Unfortunately, using the step size governed by the small scale structure of the solution evolves the large scale structure very slowly to the final answer. In principle it is obvious how to improve the algorithm: use a length scale dependent step size. It is the practice which is hard.

A technique called Fourier acceleration has been used to speed up the convergence of gauge fixing and the calculation of quark propagators in QCD [1]. In this method, a length scale dependent step size is implemented, using a fast Fourier transform to allow the step size to be a function of momentum. The functional form of the step size is chosen by studying the free field abelian theory.

However, an individual configuration from a simulation is not translationally invariant, nor "smooth", so it is not clear that the theoretical gains of Fourier acceleration will be achieved in practice. The initial work of the Cornell group, using Fourier accelerated algorithms to simulate QCD on $8^4$ lattices, showed big savings in computer time. However a later study by Hockney [2] on the calculation of quark propagators on lattice volumes between $8^4$ and $16^4$ failed to show the expected and crucial volume dependence of Fourier acceleration. In this paper, we introduce an acceleration method, similar in spirit to Fourier acceleration, based on the wavelet transform.

## 2. INTRODUCTION TO WAVELETS

Wavelets are a new field in harmonic analysis [3,4], which provide a synthesis of ideas from various fields including rigorous renormalisation group studies [5]. We shall only quote the relevant results for algorithms.

Wavelets are a set of basis functions for square integrable functions, $L^2(R)$, which unlike the exponential factors of the Fourier transform are simultaneously localised in position and momentum space. A function $f(x) \in L^2(R)$ has the following expansion in terms of the wavelet basis $\psi_{ij}(x)$

$$f(x) = \sum_{i=-\infty}^{\infty} \sum_{j=-\infty}^{\infty} c_{ij} \psi_{ij}(x) \qquad (1)$$

The $\psi_{ij}(x)$ are generated by translations and dilations of a single function $\psi(x)$

$$\psi_{ij}(x) = 2^{\frac{i}{2}} \psi(2^i x - j) \qquad (2)$$

The wavelet basis functions $\psi_{ij}(x)$ are orthogonal over different length scales, which motivates their use in fighting critical slowing down.

$$\langle \psi_{ij}, \psi_{lm} \rangle = \delta_{il} \delta_{jm} \qquad (3)$$

The practical utility of the almost magical equation 3 rests on the choice of the function $\psi(x)$. In this work we used the wavelet family due to Daubechies [3]. All the DAUB wavelets have compact support and are characterised by the vanishing of the moments of $\psi(x)$

$$\int \psi(x) x^n dx = 0, \ n = 1, \cdots \quad (4)$$

As $n$ increases the wavelets get bigger and "smoother". The DAUB2 wavelet transform is the Haar transform. We remark that wavelets can be used to characterise smoothness spaces [3] (e.g. Sobolev spaces), and this could be useful for the development of fast algorithms. The folk lore in Fourier acceleration [1] is that the practical speed of a QCD algorithm depends on how "smooth" the underlying gauge configuration is.

Discrete wavelet transforms exist, with fast algorithms and published codes [6]. A Daubechies wavelet transform will transform a vector of length $N$ in $O(N)$ operations. We used tensor products of one dimensional wavelet transforms for the three dimensional transform. More sophisticated multidimensional wavelet transforms exist [3], and look potentially useful.

## 3. COULOMB GAUGE FIXING

The gauge transformation G(x) which fixes a gauge configuration $U_\mu(x)$ into Coulomb gauge is found by maximising the following function

$$F = \frac{1}{2n_c 3V} \sum_x \sum_{k=1}^{3} \text{trace}(U_k^g(x) + U_k^g(x)^\dagger) \quad (5)$$

where the gauge transformed configuration is

$$U_k^g(x) = G(x) U_k(x) G^\dagger(x + \hat{k}) \quad (6)$$

The algorithm used is a steepest descent method [1]. The gauge transformation for a single iteration is

$$G(x) = \exp(i w_a(x) T_a \alpha) \quad (7)$$

where $T_a$ are the generators of $SU(3)$, $\alpha$ is a step size parameter, and the function $w_a(x)$ is defined by

$$w_a(x) = -i \sum_{k=1}^{3} \text{trace } T_a \Delta_{-k}(U_k(x) - U_k^\dagger(x)) \quad (8)$$

The covariant derivative $\Delta_{-k}$ is defined by

$$\Delta_{-k} U_k(x) = U_k(x - \hat{k}) - U_k(x) \quad (9)$$

We monitor the convergence of the algorithm, by calculating the lattice analog of div A.

$$\text{div A} = \frac{1}{V} \sum_{\underline{x}} \frac{1}{2} \sum_a w_a(x)^2 \quad (10)$$

## 4. WAVELET ACCELERATION

The wavelet preconditioned gauge transform is

$$G(x)_{\hat{D}_r} = \exp(\hat{D}_r^{-1} P \hat{D}_r \ i w_a(x) T_a \alpha) \quad (11)$$

where $\hat{D}_r$ is a wavelet transform labelled by r and P is a diagonal preconditioner matrix. This should be compared to the Fourier accelerated gauge transform [1]

$$G(x)_{FFT} = \exp(\hat{F}^{-1} \frac{p_{max}^2}{p^2} \hat{F} \ i w_a(x) T_a \alpha) \quad (12)$$

where $\hat{F}$ is the fast Fourier transform.

In this exploratory study, we used the simplest form of preconditioner for the wavelet accelerated gauge transformation [7]. If the dimension of the space is $2^n$, where $n$ is an integer, the preconditioner matrix P in one dimension is defined as

$$\begin{aligned} P_{ij} &= \delta_{ij} \mathcal{P}(i) \\ \mathcal{P}(i) &= 2^{n-q} \ , 2^{q-1} < i \leq 2^q \ , q = 1 \cdots n \\ \mathcal{P}(1) &= 2^{n-1} \end{aligned} \quad (13)$$

This corresponds to weighting the smallest scale structure with the smallest weight. The next biggest structure is weighted with twice the weight of the smallest structure, and so on, up the scales in the problem.

Our wavelet transform subroutines are based on the routines in Numerical Recipes [6], which stop the recursive algorithm at the second largest length scale.

## 5. NUMERICAL RESULTS

We compared the performance of the different gauge fixing algorithms on $8^4$, $\beta = 5.7$, quenched $SU(3)$ gauge configurations, generated by the pseudo heat bath algorithm. The calculations were performed at Lawrence Livermore National Lab, using a Cray YMP-C90 for the Fourier acceleration and a Cray-2 to test all the other algorithms. We do not present timings of the programs because we have not fully optimized our codes.

For each algorithm we found the maximum value of the step size parameter $\alpha$ which could be used without destabilizing the algorithm. Figure 1 shows div A as a function of iteration number. This was similiar to all the runs we did, except sometime the DAUB2 acceleration was a bit better than the unaccelerated algorithm. The practical performance of each algorithm is a function of time slice. For example, on one time slice we tested, the DAUB6 gauge fixing algorithm only reduced the number of iterations by a factor of 1.2 over the unaccelerated algorithm; for comparison, on the same time slice, Fourier acceleration reduced the number of iterations by a factor of 3.4. For the wavelet transforms we tested, the acceleration improves as the wavelet gets smoother. The step size parameter for the unaccelerated al-gorithm ($\alpha = 0.19$) was an order of magnitude larger than for the DAUB6 wavelet algorithm ($\alpha = 0.015$). This could be caused by the preconditioner we used having a destabilizing large step size for the biggest length scale. We also tried using a preconditioner which increased by a factor of $\beta$ between length scales, where $\beta$ lies between 1 and 2. By decreasing $\beta$ from 2 and increasing $\alpha$ in an ad-hoc way, we were able to improve the performance of the wavelet algorithm. This suggests that there are gains to be made by systematically tuning the parameters of a more sophisticated preconditioner.

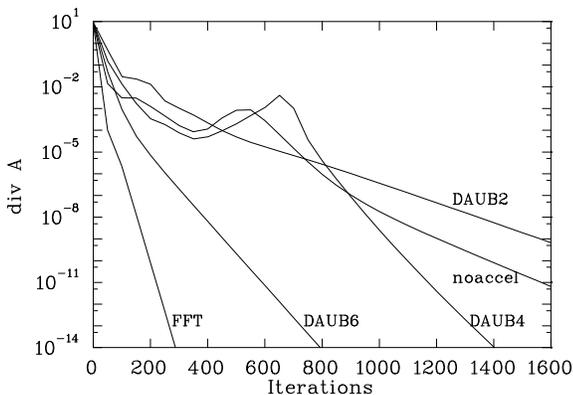

Figure 1. Comparison of various transform accelerated gauge fixing algorithms.

## 6. CONCLUSIONS

Although for the $8^4$ lattices of this study, Fourier acceleration is superior to wavelet acceleration, we have indicated the areas where the wavelet algorithms could be improved. A clearer picture will emerge after we have studied the volume dependence of all the gauge fixing algorithms.

## ACKNOWLEDGMENTS

We would like to thank Peter Perry for introducing us to wavelets. This work is supported in part by the U.S. Department of Energy under grant number DE-FG05-84ER40154 and by the National Science Foundation under grant number STI-9108764.